\documentstyle[12pt]{article}

\def\dfrac{\displaystyle\frac}
\def\i{\imath}
\def\d{\partial}
\def\noi{\noindent}
\def\pp{p_\parallel}
\def\ve{\varepsilon}
\def\bs{\bigskip}

\newcommand{\Eq}[1]{Eq.(\ref{#1})}

\newcommand{\refc}[1]{Ref.~\cite{#1}}

\newcommand{\bea}{\begin{eqnarray}}
\newcommand{\eea}{\end{eqnarray}}
\newcommand{\be}{\begin{equation}}
\newcommand{\ee}{\end{equation}}
\newcommand{\bc}{\begin{center}}
\newcommand{\ec}{\end{center}}
\newcommand{\ba}{\begin{array}}
\newcommand{\ea}{\end{array}}

\newcommand{\cL}{{\cal L}}

\newcommand{\annp}[3]{{\it  Ann. Phys. (N.Y.) }{{\bf #1} {(#2)} {#3}}}

\newcommand{\fort}[3]{{\it Fortsch. Phys. }{ {\bf #1} {(#2)} {#3}}}

\newcommand{\np}[3]{{\it  Nucl. Phys. }{{\bf #1} {(#2)} {#3}}}
\newcommand{\pr}[3]{{\it Phys. Rev.}{{ \bf #1} {(#2)} {#3}}}

\newcommand{\prl}[3]{ {\it Phys. Rev. Lett.}{{ \bf #1} {(#2)} {#3}}}

\normalsize

\begin{document}

\large
\thispagestyle{empty}
\begin{flushright}                              FIAN/TD/94-10\\
                                                hep-ph/9412204\\
                                                November 1994

\vspace{3ex}
\end{flushright}
\bc
\normalsize
{\LARGE\bf Low-Temperature  QED with External Magnetic Field}

\vspace{3ex}

{\Large Vadim Zeitlin$^{\dagger}$}

{\large Department of Theoretical Physics, P.~N.~Lebedev Physical
Institute,

  Leninsky prospect 53, 117924 Moscow, Russia}\vspace{5ex}

\ec

\centerline{{\Large\bf Abstract}}

\normalsize

\begin{quote}

Low-temperature expansion of the effective Lagrangian of the  QED$_{3+1}$ with
a uniform magnetic field and a finite chemical potential is performed.
Temperature corrections, as well as zero-temperature expression for the
effective Lagrangian are presented as finite sums over partially filled Landau
levels.

\end{quote}

\vfill
\noindent
$^\dagger$ E-mail address: zeitlin@lpi.ac.ru

\newpage


\setcounter{page}{1}

\section{Introduction}

Different approaches [1-4] may be used in order to calculate one-loop effective
Lagrangian of the finite temperature and density quantum electrodynamics with a
uniform magnetic field. The general expressions for the effective Lagrangian
obtained are rather complicated and serve, usually, only as a starting point
for subsequent approximations or direct numerical calculations.

In \refc{PerZeit} it was shown that for a particular case
$T=0,~\mu \ne 0$ one may  present an exact expression for the effective
Lagrangian
$\cL^{{\rm eff}} (T=0, B, \mu)$ in terms of the elementary functions as a
finite
sum over partially filled Landau levels. In the above-mentioned paper the
effective
Lagrangian was calculated in two ways: by using the proper-time method at zero
temperature \cite{ChodosEO90} and by taking a zero-temperature limit of
$\cL^{{\rm eff}} (B,
\mu, T)$. Here we shall show that for $T^2 \ll eB, ~~T^2 \ll \mu^2 - m^2$~ it
is
possible to perform a low-temperature expansion of the effective Lagrangian
keeping finite number of (partially) filled Landau levels and to get a similar
 representation for the temperature corrections.  Using the expression for the
 effective Lagrangian we shall obtain temperature corrections to the fermion
density and de Haas -- van Alphen oscillations.  On the base of the expression
for the fermion density we shall also calculate some components of the one-loop
polarization operator in the static limit as well as Hall conductivity in the
 QED$_{3+1}$.

\bs
We shall  consider the finite density  QED with a uniform
magnetic field.  In the presence of the chemical potential
the corresponding Lagrangian is:

\be
	\cL = -\frac14 F_{\mu\nu}F^{\mu\nu} +\bar{\psi}(i {\partial
         \kern-0.5em/} -e {A\kern-0.5em/}  - \gamma_0 \mu -m)\psi~~~.
\label{tree}
\ee

\bs
\noi
We choose here the  external magnetic field to be  parallel to the $z$--axis,
$F_{12}=-F_{21}=B$.

The effective Lagrangian at
$T,~\mu,~B \ne 0$, ~$\cL^{{\rm eff}} (B, \mu, T)$
may be written as follows [2-4]:

        \be
        \cL^{{\rm eff}} (B, \mu, T) =   \cL^{\rm eff}(B)+
        \tilde\cL^{\rm eff}(B,\mu,T)~~~,
        \ee

\bs
\noi
where $\tilde\cL^{\rm eff}(B,\mu,T)$,

        \bea
        \tilde\cL^{\rm eff}(B,\mu,T)=
        \frac1{\beta}\frac{eB}{(2\pi)^2} \sum_{k=0}^\infty
        b_k \int_{-\infty}^\infty d\!p_\parallel \left\{
        \ln[ 1+e^{-\beta(\varepsilon_k(p_\parallel)-\mu)}] +
        \ln[ 1+e^{-\beta(\varepsilon_k(p_\parallel)+\mu)}] \right\}~~~.
        \label{lagr1}
\eea

\bs
\noi
is  the contribution due to the finite temperature and density
($\beta= 1/T$, ~$p_\parallel$ is the modulus of the momentum parallel to the
magnetic field, $b_k\equiv 2- \delta_{n,0}~$,
since the lowest Landau level ($n=0$), unlike the higher ones,
contains the fermions with only one spin projection),
and  $\cL^{\rm eff}(B)$,

        \be
        \cL^{{\rm eff}} (B)=-\frac{1}{8\pi^2} \int_0^\infty \frac{ds}{s^3}
        \left[ eBs \coth(eBs)-1-\frac13 (eBs)^2 \right]
        \exp(-m^2s)
        \ee

\bs
\noi
is the Schwinger  Lagrangian in  the purely magnetic case
\cite{Schwinger51}.

\bigskip
\section{Low-temperature expansion at $\mu,B \ne 0$}
Integrating  \Eq{lagr1} by parts one gets:

        \be
        \tilde\cL^{\rm eff}(B,\mu,T)=
        \frac{eB}{(2\pi)^2} \sum_{n=0}^\infty
        b_n \int_{-\infty}^\infty d\! \pp ~\frac{\pp^2}{\ve_n(\pp)}
        ~(f_+(T) + f_-(T))~~~,
        \label{lagr2}
        \ee

\bs
\noi
$f_\pm(T)$ denotes the Fermi distribution,

        \be
        f_\pm(T) = \frac1{1+e^{\beta(\ve\mp \mu)}} \quad .
        \nonumber
        \ee

\bs
We shall consider here the low-temperature limit of the QED$_{3+1}$.
In the $T \rightarrow 0$ limit the Fermi distribution approaches the
step-function, $\lim_{T \rightarrow 0} f_\pm = \theta(\pm\mu - \ve)$ and
\Eq{lagr2} reads  \cite{PerZeit}:

        \bea
        \lefteqn{\tilde\cL^{\rm eff}(T = 0,B,\mu)=}\nonumber\\
        &&\label{zeroT}\\
        &&  \frac{eB}{(2\pi)^2}
        \sum_{n=0}^{ \left[   \frac{\mu^2 - m^2}{2eB}   \right]}
        b_n
        \left\{ \mu
        \sqrt{\mu^2-m^2-2eBn}
        - (m^2+2eBn)
        \ln     \left(
        \frac{ \mu + \sqrt{\mu^2-m^2-2eBn}} {\sqrt{m^2+2eBn}}
                                              \right) \right\}\nonumber
        \eea

\bs
\noi
where $[ \dots ]$ denotes the integral part.

\bs
To evaluate the low-temperature corrections to the effective Lagrangian
(\ref{lagr2}) we shall calculate a derivative of $\tilde\cL^{\rm eff}(B,\mu,T)$
with respect to $T$ first:

        \be
        \frac{\d \tilde\cL^{\rm eff}(B,\mu,T)}{\d T}=
        \frac{eB}{(2\pi)^2} \sum_{n=0}^\infty
        b_n \int_{-\infty}^\infty d\!\pp \frac{\pp^2}{\ve_n(\pp)}
        \left(
        \frac{\d f_+}{\d T} +\frac{\d f_-}{\d T}
                                                        \right)~~~.
        \ee

\bs
The derivatives of the Fermi distributions may be rewritten in the following
way\footnote{We shall consider a system with a fixed chemical potential,
and not a density, $\dfrac{\d \mu}{\d T} = 0$, cf. calculation of the heat
capacity in \refc{Abrikosov}.}:

        \be
        \frac{\d f_\pm}{\d T} =
	- \frac{\ve \mp \mu}{T}
        ~\frac{\d f_\pm}{\d \ve}~~~,
        \ee

\bs
\noi
and

        \be
        \frac{\d \tilde\cL^{\rm eff}(B,\mu,T)}{\d T}=
        -\frac{eB}{(2\pi)^2} \sum_{n=0}^\infty
        b_n \int_{-\infty}^\infty d\!\pp \frac{\pp^2}{\ve_n(\pp)}
        \left(
	\frac{\ve_n - \mu}{T}
        ~\frac{\d f_+}{\d \ve_n} +
	\frac{\ve_n + \mu}{T}
        ~\frac{\d f_-}{\d \ve_n}
        \right)~~~.
        \label{dT1}
        \ee

\bs
Making change of variables in \Eq{dT1}  one has:

        \bea
        \lefteqn{\frac{\d \tilde\cL^{\rm eff}(B,\mu,T)}{\d T}=}\nonumber\\
        &&\label{dT2}\\
        &&-\frac{eB}{(2\pi)^2} \sum_{n=0}^\infty
        b_n \int_{(m^2+2eBn)^{1/2}}^\infty d\!q\/
                \left(
	\frac{q - \mu}{T}
        ~\frac{\d f_+}{\d q}
	+
	\frac{q + \mu}{T}
        ~\frac{\d f_-}{\d q}
                                        \right)
        \sqrt{q^2 - m^2 - 2eBn} ~~~. \nonumber
        \eea

\bs
The derivatives $\dfrac{\d f_\pm}{\d q}$ may be rewritten as follows:

        \be
        \frac{\d f_\pm}{\d q} = - \frac1{4T} \cdot
        \frac1{\cosh^2((q\mp\mu)/2T)}~~~.
        \ee

\bs
\noi
The above functions decrease sharply for small $T$ as one
moves off the point $q=\pm\mu$ (as
$f_\pm\rightarrow\theta(\pm\mu-\ve)$ the derivative of the Fermi
distribution with respect to energy approaches the  $\delta$-function).
Therefore, only partially filled Landau
levels (i.e. those with their edge situated below the Fermi surface, $\mu^2 >
m^2
+ 2eBn$)
will contribute to the effective Lagrangian in a low-temperature limit.

Now we may introduce an auxiliary variable $z = q \mp \mu$ with the integration
limits \nolinebreak $\pm\infty$ and rewrite \Eq{dT2} as follows ($\mu>0$):

        \be
        \frac{\d \tilde\cL^{\rm eff}(B,\mu,T)}{\d T}=
        \frac{eB}{(2\pi)^2}
        \sum_{n=0}^{\left[   \frac{\mu^2 - m^2}{2eB}   \right]}
        b_n \int_{-\infty}^\infty d\!z
        \frac{z}{4T^2} \frac1{\cosh^2
        \left(          \frac{z}{2T}                    \right)}
        \sqrt{(z+\mu)^2 - m^2 - 2eBn}~~~.
        \ee

\bs
Expanding $\sqrt{(z+\mu)^2 - m^2 - 2eBn}$ ~at $z=0$ and keeping a
leading term only (the even powers of $z$ will not contribute to $\dfrac{\d
\cL}{\d T}$) one finally has:

        \bea
        \lefteqn{\frac{\d \tilde\cL^{\rm eff}(B,\mu,T)}{\d T} =}\nonumber\\
        & &\nonumber\\
        & &
        \frac{eB}{(2\pi)^2}
        \sum_{n=0}^{ \left[   \frac{\mu^2 - m^2}{2eB}   \right]}
        b_n \int_{-\infty}^\infty d\!z
        \frac{z^2}{4T^2}\frac1{\cosh^{2}( \frac{z}{2T})}
        \frac{\mu}{(\mu^2 - m^2 - 2eBn)^{1/2}} + O(T^3) =\nonumber \\
        & &\nonumber\\
        & &
        \frac{eBT}6
        \sum_{n=0}^{ \left[   \frac{\mu^2 - m^2}{2eB}   \right]}
        b_n \frac{\mu}{(\mu^2 - m^2 - 2eBn)^{1/2}} + O(T^3)~~~.
        \eea

\bs
The above expansion is valid at

	$$
        \frac{T}{(\mu^2 - m^2 - 2eBn)^{1/2}} \ll 1~~~,
	$$

\bs
\noi
i.e. as long as the distance from the Fermi
surface to the edge of any Landau level remains much greater than the
temperature.

Now we can write a low-temperature expansion for the one-loop effective
Lagrangian:

        \bea
        \lefteqn{\tilde\cL^{\rm eff}(T,B,\mu)=}\nonumber\\
        & &\nonumber\\
        &&  \frac{eB}{(2\pi)^2}
        \sum_{n=0}^{ \left[   \frac{\mu^2 - m^2}{2eB}   \right]}
        b_n \left\{ \mu
        \sqrt{\mu^2-m^2-2eBn}
        - (m^2+2eBn)
        \ln     \left(
        \frac{ \mu + \sqrt{\mu^2-m^2-2eBn}} {\sqrt{m^2+2eBn}}
                                              \right) \right.\nonumber\\
        & &\nonumber\\
        && {+ \left.
        \frac{T^2\pi^2}3 \frac{\mu}{(\mu^2 - m^2 - 2eBn)^{1/2}}
        \right\}
        + O(T^4)}
        \eea

\bs
\noi
as well as for the fermion density, $\rho = \dfrac{\d \cL}{\d \mu}$:

        \bea
        \lefteqn{\rho(B,\mu,T) =}\nonumber\\
        &&\label{rho}\\
        &&
        \frac{eB}{2\pi^2}
        \sum_{n=0}^{\left[  \frac{\mu^2 - m^2}{2eB} \right]}
        b_n \sqrt{\mu^2-m^2-2eBn}
        \left\{
        1 - \frac{T^2\pi^2}6
        \frac{m^2 + 2eBn}{(\mu^2-m^2-2eBn)^2}      \right\} + O(T^4)~~~.
        \nonumber
        \eea

\bs
\noi
Thermal corrections decrease the fermion density:
by fixing the chemical potential and raising the
temperature we evaporate the electron (positron) gas.

\bigskip
\section{Fermion density and polarization operator}

Having the expressions for the effective Lagrangian and the fermion density we
may move
forward to calculate magnetization, Hall conductivity and some
components of the polarization operator in the static limit ~~$p_0=0, ~~{\bf p}
\rightarrow 0$.

The magnetization $M$ is defined as
$M=\dfrac{\partial \cL^{\rm eff}}{\partial B}$. Here $M(B,\mu,T) =
M(B) + \tilde{M}(B,\mu,T)$, $M(B)$ is the vacuum magnetization,  and $\tilde
M=\dfrac{\partial \tilde\cL^{\rm eff}}{\partial B}$

        \bea
        \lefteqn{\tilde{M}(B,\mu,T)=}\nonumber\\
        &&\nonumber\\
        &&
        \frac{e}{4\pi^2}
        \sum_{n=0}^{\left[  \frac{\mu^2 - m^2}{2eB} \right]}
        b_n
        \left\{
        \mu    \sqrt{\mu^2-m^2-2eBn}
         - (m^2+4eBn) \ln
        \left( \frac{ \mu +
        \sqrt{\mu^2-m^2-2eBn}}
        {\sqrt{m^2+2eBn}}
                        \right)  \right.\nonumber\\
        &&\nonumber\\
        &&
        \left.
        + \frac{\pi^2T^2\mu (\mu^2 - m^2 -
        eBn)}{(\mu^2 - m^2 - 2eBn)^{3/2}} \right\}~~~.
        \eea

\bs
\noi
i.e. the temperature corrections smoothing de Haas--van Alphen oscillations
(cf.
\refc{ElmforsPS93,PerZeit}).

\bs
The $\Pi_{00}$--component of the polarization operator
may be written in the static limit as a derivative of
density with respect to chemical potential \cite{Fradkin},
{}~$\Pi_{00}(p_0=0,~{\bf p}\rightarrow 0) = e^2 \dfrac{\d \rho}{\d \mu}$

        \bea
        \lefteqn{\Pi_{00}(p_0=0,~{\bf p}\rightarrow 0)=}\nonumber\\
        &&\\
        &&e^2
        \frac{eB\mu}{2\pi^2}
        \sum_{n=0}^{\left[  \frac{\mu^2 - m^2}{2eB} \right]}
        b_n
        \left\{
        (\mu^2-m^2-2eBn)^{- 1/2}
        + T^2\pi^2
        \frac{m^2 + 2eBn}{(\mu^2-m^2-2eBn)^{5/2}}         \right\}~~~.
        \nonumber
        \eea

\bs
At $B=0$ ~~$\Pi_{00}(p_0=0,~{\bf p}\rightarrow 0)$ defines the Debye's radius,
$\lambda^{-2} =\Pi_{00}(p_0=0,~{\bf p}\rightarrow 0)$ \cite{Fradkin} but this
relation does not hold for $B\ne0$ as the tensor structure of the polarization
operator is now more complicated (see below).

\bs
The components $\Pi_{01}=\Pi_{10}^*$ and $\Pi_{02}=\Pi_{20}^*$  in the static
limit may be written as

        \be
        \Pi_{0j}(p \rightarrow 0) = \i e\/ \varepsilon_{ij}p_i\frac{\d
        \rho}{\d B} \quad i,j = 1,2~~~,
        \ee

\bs
\noi
which follows from the definition of the polarization operator,

        \be
        \Pi_{\mu\nu}(x,x') =
        \i~ \frac{\delta <j_\mu (x)>}{\delta A_\nu (x')}~~~.
        \ee

\bs
The above components of the polarization operator describe conductivity in
the plane orthogonal to the magnetic field. Indeed, a
current induced by a perturbating electric field in the presence of a
strong magnetic field may be written as a
linear response function,

        \be
        \tilde{I}_\mu(x) = \int d^4\! x' \Pi_{\mu\nu}(x - x'|B,\mu)
        A_{pert}^\nu(x')~~~.
        \ee

\bs
Choosing $A_{pert}^\nu(x)$ as $A_{pert}^0(x) = {\bf x}\cdot{\bf  E},
{}~{\bf A}_{pert}(x) = 0 $ one has the following expression for the
conductivity
$\sigma_{ij}$:

        \be
        \sigma_{ij} =
        \left. \frac{\d j_i}{\d E_j} \right|_{E \rightarrow 0}=
        \i \left. \frac{\d \Pi_{0i}(p)}{\d p_j}
                \right|_{p \rightarrow 0}~~~.
        \ee

\bs
Thus we can see that the conductivity in the plane orthogonal to the magnetic
field is Hall-like,

        \be
        \sigma_{ij} = \varepsilon_{ij} \sigma^{Hall}
        = e \varepsilon_{ij} \frac{\d \rho}{\d B} \quad i,j = 1,2 ~~~.
        \label{sigma}
        \ee

\bs
Before examining this formal expression it is worth looking at the
general structure of the polarization
operator in order to find an element responsible for the Hall
conductivity. The polarization tensor in QED$_{3+1}$ with a uniform magnetic
field at $T,\mu \ne 0$ may be decomposed over six tensor structures
\cite{PRShabad79}:

        \bea
        \lefteqn{\Pi_{\mu\nu}(p|T,\mu,B)=}\nonumber\\
        &&\nonumber\\
        &&\left(
        g_{\mu\nu} - \frac{p_\mu p_\nu}{p^2}               \right){\cal A}
        +\left(
        \frac{p_\mu p_\nu}{p^2} -
        \frac{p_\mu u_\nu + u_\mu p_\nu}{(pu)} +
        \frac{u_\mu u_\nu}{(pu)^2} p^2                          \right)
        {\cal B} + \nonumber\\
        &&\nonumber\\
        && \left(
        g_{\mu\lambda} - \frac{p_\mu p_\lambda}{p^2}    \right)
        F^{\lambda\rho}F_\rho^\phi
        \left(
        g_{\nu\phi} - \frac{p_\nu p_\phi}{p^2}
        \right){\cal C}
        + F_{\mu\lambda}p^\lambda F_{\nu\phi}p^\phi {\cal D} +\\
        &&\nonumber\\
        &&  \i \left(
        p_\mu F_{\nu\lambda}p^\lambda  -
        p_\nu F_{\mu\lambda}p^\lambda  +
        p^2F_{\mu\nu}                   \right) {\cal E}
        + \i \left(
        u_\mu F_{\nu\lambda}p^\lambda  -
        u_\nu F_{\mu\lambda}p^\lambda  +
        (pu)F_{\mu\nu}                   \right) {\cal F}~~~.\nonumber
        \eea

\bs
The scalar functions ${\cal A, B, C, D, E}$ and ${\cal F}$ being the functions
of
$p_0^2, ~p_3^2, ~p_1^2 + p_2^2$ and $B$
($u^\mu$ is the $4$-velocity of the medium \cite{Fradkin}, $u^\mu = (1,0,0,0)$)
, only the last tensor
structure may contribute to the Hall conductivity
\footnote{For $B=0$, $T$ and/or $\mu \ne
0$ ~~${\cal C,D,E, F} \equiv 0$, for $B,T \ne 0, ~\mu = 0$ ~~${\cal E, F}
\equiv 0$, for $T,\mu = 0, ~B\ne 0$ ~~${\cal B, E, F} \equiv 0$, see
\refc{Shabad88} for details.}. Therefore, we may
define the coefficient ${\cal F}$ in the static limit as well:

        \be
        {\cal F}(p_0=0,~{\bf p}\rightarrow 0) = \frac{e}{B}
        \frac{\d \rho}{\d B}~~~.
        \ee

\bs
Substituting into \Eq{sigma} the expression for the fermion density
\Eq{rho}
one has (we are keeping only $T=0$ terms here for simplicity):

        \be
        \sigma^{Hall}(B,\mu) =   \frac{e}{2\pi^2}
        \sum_{n=0}^{\left[  \frac{\mu^2 - m^2}{2eB} \right]}
        b_n\frac{\mu^2-m^2-3eBn}{(\mu^2-m^2-2eBn)^{1/2}}~~~.
        \label{hall}
        \ee

\bs
It follows from \Eq{hall} that the Hall conductivity in the QED$_{3+1}$ is
an oscillating function (of infinite amplitude) of the chemical potential and
the magnetic field.
The inverse square-root singularity is not
surprising as the polarization operator in QED$_{3+1}$ with a uniform
magnetic field at $T = 0$ has just the same kind of singularities
\cite{PRShabad82} and $\sigma^{Hall}$ is proportional to one of its scalar
coefficients. On the other hand, it follows from
\Eq{hall} that the sign of conductivity is changing as one moves from the edge
of one Landau level to another. At $T \ne 0$ the conductivity is a smooth
function of $B,\mu$  but the density is not a monotonous function of $B$ (at
least at low temperatures), therefore, the sign of conductivity will still
depend on the magnitude of the magnetic field.

This will hold for the nonrelativistic case as well: in
the nonrelativistic limit one has just to change $\mu^2 -m^2$ to  $2m\mu^*$
in \Eq{hall}. At the same time, there is no evidence of such dependence
of conductivity in realistic condensed matter systems. There may be two
possible
explanations of this fact.  First, we have considered a nondissipative plasma.
In
 a real condensed matter system $\dfrac{\mu^2 - m^2}{2eB} = \dfrac{m\mu^*}{eB}
\gg 1$.  In this case even small widths of levels (a finite free-path length)
may
cancel the oscillations completely. Second, we have considered the one-loop
approximation.  Higher order corrections may also level the  oscillations.

\bigskip
\section*{Acknowledgement}
I am  grateful to V.~V.~Losyakov for fruitful discussions.
This work supported in part by the Russian Foundation for  Basic Research
Grant $N^o$ 67123016 and Grant MQM000 from the International Science
Foundation.


\newpage

\bigskip
\begin{thebibliography}{99}

\bibitem{Canuto68}
                V.~Canuto and H.-Y.~Chiu, \prl{21}{1968}{110};\\
                V.~Canuto and H.-Y.~Chiu,
                \pr{173}{1968}{1210}.

\bibitem{PRShabad76}
                H.~P\'erez Rojas and A.~E.~Shabad,
                {\it Bull. Lebedev Phys. Inst.} {\bf 7} (1976) 16.


\bibitem{Cabo81}
                A.~Cabo,
                \fort{29}{1981}{495}.

\bibitem{ElmforsPS93}
                P.~Elmfors, D.~Persson, and  B.-S.~Skagerstam,
                \prl{71}{1993}{480};\\
                P.~Elmfors, D.~Persson, and  B.-S.~Skagerstam,
                {\it Astroparticle Phys.\/} {\bf 2} (1994) 299.

\bibitem{PerZeit}
                D.~Persson and Vad.~Zeitlin,
                preprint ( FIAN/TD/94-01, G\"oteborg ITP 93-11,\\
                hep-ph/9404216),
                {\it Phys.~Rev.} {\bf D} (in press).

\bibitem{ChodosEO90}
                A. Chodos, K. Everding, and D. A. Owen,
                \pr{D42}{1990}{2881}.

\bibitem{Schwinger51}
                J.~Schwinger, \pr{82}{1951}{664}.


\bibitem{Abrikosov}
                A.~A.~Abrikosov, {\it Fundamentals of the Theory of Metals},
                North-Holland, 1988.

\bibitem{Fradkin}
                E.~S.~Fradkin, \np{12}{1959}{465};\\
                E.~S.~Fradkin, {\it Proc. Lebedev Phys. Inst.}
                {\bf 29} (1965) 7.


\bibitem{PRShabad79}
                H.~P\'erez Rojas and A.~E.~Shabad, \annp{121}{1979}{432}.

\bibitem{Shabad88}
                A.~E.~Shabad, {\it Proc. Lebedev Phys. Inst.}
                {\bf 192} (1988) 5.

\bibitem{PRShabad82}
                 A.~E.~Shabad, \annp{90}{1975}{160}.


\end{thebibliography}
\end{document}